\documentclass[prb,twocolumn,showpacs,preprintnumbers,amsmath,amssymb]{revtex4}
\usepackage[dvips]{graphicx}
\usepackage{dcolumn}
\usepackage{bm}
\usepackage{times}
\usepackage{mathptmx}
\usepackage{color}
\usepackage{pifont}
\usepackage[latin1]{inputenc}

\begin{document}

\title{Interfacial Coupling in Multiferroic-Ferromagnet Heterostructures}

\author{M. Trassin,$^{1,2,*}$ J. D. Clarkson,$^3$ S. R. Bowden,$^{4,5}$ Jian Liu,$^1$ J. T. Heron,$^3$ R. J. Paull,$^3$ E. Arenholz,$^6$ D. T. Pierce,$^4$ and J. Unguris$^4$}

 \affiliation{$^1$Department of Physics, University of California, Berkeley, CA 94720, USA}
 \affiliation{$^2$Department of Materials, ETH Zurich, Wolfgang-Pauli-Strasse 10, 8093 Zurich, Switzerland}
 \affiliation{$^3$Department of Materials Science and Engineering, University of California, Berkeley, CA 94720, USA}
 \affiliation{$^4$Center for Nanoscale Science and Technology, National Institute of Standards and Technology, Gaithersburg, MD 20899-6202, USA}
 \affiliation{$^5$Maryland Nanocenter, University of Maryland, College Park, MD 20742}
 \affiliation{$^6$Advanced Light Source, Lawrence Berkeley National Laboratory, Berkeley, California 94720, USA}
\date{\today}

\begin{abstract}
We report local probe investigations of the magnetic interaction between BiFeO$_3$ films and a ferromagnetic Co$_{0.9}$Fe$_{0.1}$ layer. Within the constraints of intralayer exchange coupling in the Co$_{0.9}$Fe$_{0.1}$, the multiferroic imprint in the ferromagnet results in a collinear arrangement of the local magnetization and the in-plane BiFeO$_3$ ferroelectric polarization. The magnetic anisotropy is uniaxial, and an in-plane effective coupling field of order 10~mT is derived. Measurements as a function of multiferroic layer thickness show that the influence of the multiferroic layer on the magnetic layer becomes negligible for 3~nm thick BiFeO$_3$ films. We ascribe this breakdown in the exchange coupling to a weakening of the antiferromagnetic order in the ultrathin BiFeO$_3$ film based on our X-ray linear dichroism measurements. These observations are consistent with an interfacial exchange coupling between the CoFe moments and a canted antiferromagnetic moment in the BiFeO$_3$.
\end{abstract}


\maketitle

\section{Introduction}

Over the past decade, a tremendous amount of research has been directed at integrating multiferroics into magnetic devices to achieve electrical control of the magnetization vector.\cite{1,2} A prominent example is bismuth ferrite, BiFeO$_3$ (BFO), a room temperature, single phase multiferroic in which the electrical control of its ferroelectric architecture and the magnetoelectric coupling to its antiferromagnetism have been widely studied.\cite{3,4,5,6,7} A second coupling, that at the interface between the antiferromagnetism in the BFO and a ferromagnetic film, is key to achieving a functional multiferroic-ferromagnetic heterostructure. The magnetoelectric coupling provides a pathway to electrically control the antiferromagnetism in the BFO and, in turn through the interfacial coupling, the ferromagnetism in the heterostructures.\cite{5} The interfacial exchange coupling has been the subject of many investigations which have shown that the details of the interface and the lateral antiferromagnetic domain structure play an important role in determining the nature of the coupling.\cite{8,9} Nevertheless, our understanding of the interfacial coupling is incomplete.

Here, we report our investigation of the interfacial coupling in Co$_{0.9}$Fe$_{0.1}$/BiFeO$_3$ (Pt/CoFe/BFO) heterostructures. We visualize 
the imprint of the multiferroic BFO domains in the ferromagnetic CoFe layer using a combination of magnetic force microscopy (MFM), 
piezoforce microscopy (PFM) and scanning electron microscopy with polarization analysis (SEMPA). High resolution SEMPA images measure for 
the first time the magnetization direction of the CoFe within each ferroelectric domain region. The energetics of the interface coupling is 
studied with vibrating sample magnetometry (VSM) and micromagnetic modeling of the spatial variation of the magnetization measured by SEMPA. 
For thick (50~nm - 150~nm) multiferroic films, we find a strong one-to-one correlation of ferromagnetic CoFe domains and the multiferroic 
domains in BFO at the single domain scale with a pronounced in-plane uniaxial magnetic anisotropy. For a 3~nm BFO thickness, the CoFe magnetic domains are decoupled from the BFO, which is manifested in dramatic changes of the magnetic behavior. Based on X-ray linear dichroism experiments on 3~nm thick BFO films, we argue that the absence of interfacial correlation is due to a weakening of the BFO antiferromagnetic order.

In antiferromagnetic ferroelectric BFO thin films, strain is responsible for suppressing the spin cycloid present in bulk crystals.\cite{3} In the strained BFO system, the antiferromagnetic axis $\mathbf L$ is perpendicular to the polarization, which points along one of the eight $<111>$ directions. The oxygen octahedral rotations break the symmetry between the two antiferromagnetic sublattices of Fe moments allowing a Dzyaloshinskii-Moryia (DM) interaction.\cite{10} As a result, a weak ferromagnetic behavior emerges in such films due to the DM interaction.\cite{3,10} In this case, a combination of the exchange interaction and spin-orbit coupling causes a canting of the moments in the antiferromagnet producing a canted net moment $\mathbf{M}_c$ giving weak ferromagnetism.\cite{11,12} The canted moment due to the DM interaction\cite{10} is perpendicular to $\mathbf L$, and perpendicular to $\mathbf P$ as illustrated in Fig. 1(a).

\begin{figure}
\centering
\includegraphics[width=0.48\textwidth]{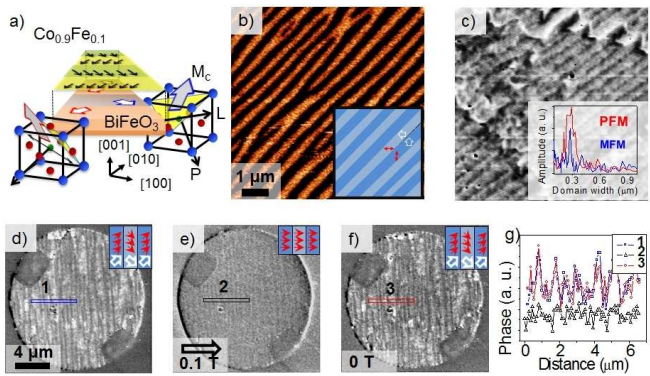}
\caption{\label{fig:fig1} (a) Schematic of coupling model showing 71$^{\circ}$ ferroelectric domains with in-plane components of $\mathbf P$ and $\mathbf M_c$ changing 90$^{\circ}$ from one domain to the next and the corresponding change in the magnetization in-plane in the CoFe layer. The cube on the right shows an example of the relative orienations of $\mathbf P$, $\mathbf M_c$, and the $\mathbf L$ axis, $[1,-1,-1]$, $[1,-1,2]$, and $[1,1,0]$ respectively, in the DM model. (b) PFM image of the BFO thin film grown on SRO/DSO. The inset shows the schematic of the polarization (open arrows) and the canted moment easy axes (double arrows) in each single domain. (c) MFM picture from the same sample after growth of the CoFe/Pt bilayer under a 20~mT growth field. In the inset the Fourier analysis shows a common domain width of 275~nm. (d-f) Magnetic field dependent MFM images as-grown (d), under 0.1~T (e) and back to 0~T (f). The inset represents the corresponding ferroelectric polarization (open arrows) and magnetization (smaller red arrows) in each domain. (g) Profile line scan in the same area in the states (d), (e) and (f) labeled 1,2 and 3 respectively.}
\end{figure}

It is expected to be energetically costly to reverse the oxygen octahedral rotations\cite{10} so that the DM interaction gives rise to a unidirectional anisotropy when $\mathbf{M}_c$ is coupled to the magnetization of the CoFe.\cite{13} However, if the energy cost of rotating the octahedra is small, then the unidirectional anisotropy is easily reversed, and the DM interaction gives rise to an effective uniaxial anisotropy. An alternate mechanism for the formation of a canted moment exists when the BFO is coupled to the CoFe. In this case, a canting of the antiferromagnetic moments lowers the exchange energy between the ferromagnet and the antiferromagnet when the ferromagnetic magnetization is perpendicular to $\mathbf L$. This coupling, along with the strong in-plane shape anisotropy in the CoFe film, creates a uniaxial anisotropy with its axis in the sample plane and perpendicular to $\mathbf L$. The two mechanisms for canted moment formation are difficult to distinguish through a direct measurement because $\mathbf{M}_c$ is so small. However, the two mechanisms should affect the domain structure of the ferromagnet differently; canted moments driven by the DM interaction are expected, assuming sufficient energy cost to reverse the oxygen octahedral rotations, to give rise to a unidirectional anisotropy, but those driven by coupling to the ferromagnet give rise to a uniaxial anisotropy.\cite{14,15,16} Both mechanisms may be present, but the correlation that we observe between the domain structures of BFO and CoFe is consistent with a uniaxial anisotropy, and, hence, consistent with a canted moment due to the interfacial exchange coupling between the CoFe moments and the BFO, or a weak barrier to reversing the rotations of the oxygen octahedra.

\section{Experimental Details}

The ferromagnet-multiferroic heterostructures were grown by a combination of pulsed laser deposition (PLD) and DC sputtering. BFO thin films with different thicknesses (3~nm to 150~nm) were epitaxially grown by PLD on SrRuO$_3$ (SRO) buffered DyScO$_3$ (DSO) $(110)$-oriented substrates, and on SRO buffered SrTiO$_3$ (STO) $(100)$-oriented substrates. No specific treatments were performed on the substrate prior to deposition. Grown on DSO substrates, heterostructures can be engineered to maintain the in-plane strain from the orthorhombic substrate which generates well ordered $71^{\circ}$ stripe domains in the BFO with two polarization variants.\cite{17} Grown on STO substrates, BFO exhibits a four-variant ferroelectric domain pattern.\cite{18,19} The topography of all the BFO thin films was investigated by atomic force microscopy (AFM) and our analysis showed negligible influence of the film thickness on the root mean square (RMS) roughness. On STO substrates the 3~nm thick BFO film exhibits a RMS roughness of 0.44~nm and the 150~nm film shows a RMS value of 0.40~nm. On DSO the 3~nm and 150~nm thick BFO films showed a RMS value of 0.22~nm and 0.25~nm, respectively. Following the BFO growth, a non magnetostrictive\cite{20} Co$_{0.9}$Fe$_{0.1}$ amorphous alloy (2.5~nm) / Pt (2.5~nm) bilayer was deposited by DC sputtering with and without a 20~mT growth field. The crystallinity and the thicknesses of the BFO films were probed by X-ray diffraction.

\section{Results and Discussion}

The underlying ferroelectric domains of BFO grown on DSO and the CoFe magnetic domains are shown in the PFM and MFM images of Fig. 1(b) and (c) prior to and following the CoFe/Pt growth. In this periodic ferroelectric architecture, the polarization across each domain wall changes by $71^{\circ}$,\cite{21} (when projected onto the $(001)$ BFO plane as measured by PFM, these angles project to $90^{\circ}$). The in-plane component of the canted moment $\mathbf M_c$ in each single stripe\cite{22} is expected to be collinear with the in-plane component of $\mathbf P$ as shown in Fig. 1(a) and in the inset of Fig. 1(b). MFM, which is sensitive to the out-of-plane component of the magnetization and allows tracking the position of the magnetic domain walls, shows that the exchange coupled CoFe ferromagnetic domains adopt the same stripe like pattern as the BFO ferroelectric domains. A common 275~nm domain width can be extracted from a Fourier analysis (see inset in Fig. 1(c)). Upon application of a magnetic field (Fig. 1(d-g)) of 0.1~T in the plane, the domain structure is erased and the image presents a uniform contrast (Fig. 1(e)), indicative of a single domain state. Strikingly, when the magnetic field is reduced to zero (Fig. 1(f)), the original magnetic domain structure in the CoFe returns. This is clearly illustrated in the image line scans in Fig. 1(g) and demonstrates the coupling of the CoFe to the magnetic order that is present at the interface in each domain in the BFO. We note that the ferromagnetic pattern disappears when a fully strained intermediate 2~nm thick SrTiO$_3$ layer is inserted in the interface between the BFO and the CoFe,\cite{5} thus ruling out magnetostrictive coupling as has been observed in other systems.\cite{23} The ferromagnetic pattern in CoFe induced by the coupling to the BFO also disappears for thick CoFe films ($>20$~nm) indicating the importance of the interface in the coupling.
We next focus on the local direction of the magnetization vector in each of the CoFe domains, using SEMPA\cite{24} (see Fig. 2).

\begin{figure}
\centering
\includegraphics[width=0.48\textwidth]{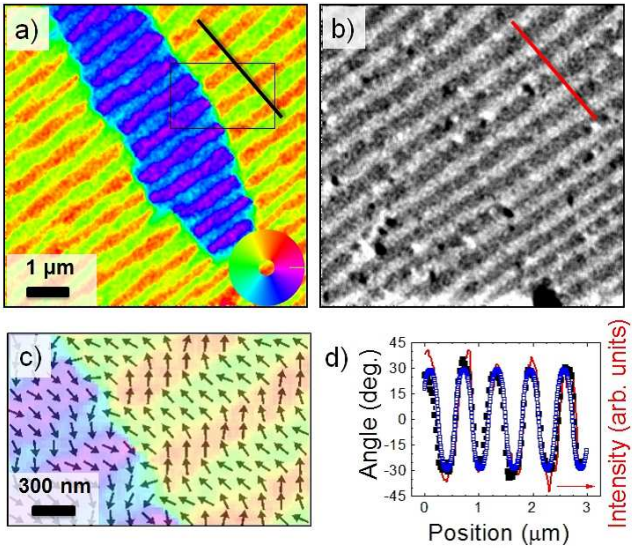}
\caption{\label{fig:fig2} (a) SEMPA image of CoFe magnetization (direction indicated by colorwheel), and (b) simultaneously acquired BSE image of underlying ferroelectric domain structure. (c) Magnified region from the SEMPA image (a) shows measured in-plane magnetization directions with arrows. (d) Line scans from BSE (red curve) and SEMPA (filled black squares) and from an OOMMF calculation (open blue squares) with a 7~mT effective coupling field showing correlated structures and the magnitude of magnetic oscillations.}
\end{figure}

This technique measures the direction of the in-plane magnetization at the surface with 20~nm resolution. (No out-of-plane magnetization was found in these samples; out-of-plane tilt $< 6^{\circ}$.) The samples were measured at remanence after cleaning and removing most of the Pt coating by ion sputtering with 800~eV Ar ions. In addition to the SEMPA measurements which use the low-energy secondary electrons, the high-energy back-scattered electron (BSE) intensity was also measured which provides a simultaneous image of the underlying ferroelectric domain structure through electron channeling contrast.\cite{25} The characteristic, stripe-like images clearly show the correlation between the ferromagnetic structure in Fig. 2(a) and the underlying ferroelectric structure in Fig. 2(b). From the SEMPA images (color wheel in inset) it can be seen that the magnetization switches between stripes while the average net magnetization is perpendicular to the stripes. Figure 2(c) shows a high resolution picture of the stripe structure in the region where the net magnetization is parallel and antiparallel to the net in-plane polarization. The presence of both magnetic alignments is an indication of the canted antiferromagnetic moment introducing a uniaxial anisotropy rather than a unidirectional anisotropy.
 
In Fig. 2(d), line scans from the same region of the SEMPA (filled black squares) and BSE (red curve) images show quantitatively the correlation between the CoFe magnetization and the BFO domains. However, whereas the symmetry restricted in-plane components of the BFO polarization and canted moment change by 90$^{\circ}$ from one stripe to the next, the CoFe magnetization changes by 60$^{\circ} ± 6^{\circ}$ ($\pm$ one standard deviation) as shown in Fig. 2(d) for the 310~nm wide stripes of this region. Micromagnetic modeling shows that this difference can be explained by the competition between the intralayer exchange energy in the CoFe film and the coupling at the interface with the BFO. We simulated this competition using the Object Oriented Micromagnetic Framework (OOMMF)\cite{26} with parameters for bulk Co. The coupling at the BFO interface is represented by an effective coupling field in the in-plane direction of the canted moment in each domain. Choosing an effective coupling field of $7$ ($+2.5$, $-1.5$)~mT provided the best fit between the angular variation of the magnetization in the simulation (open blue squares in Fig. 2 (d)) and the data.
 
An additional measure of the strength of the interfacial coupling and its dependence on BFO thickness was obtained from room temperature VSM measurements that were carried out on different heterostructure stacks, i.e.\ with and without an external magnetic growth field, and on both DSO and STO substrates.

\begin{figure}
\centering
\includegraphics[width=0.48\textwidth]{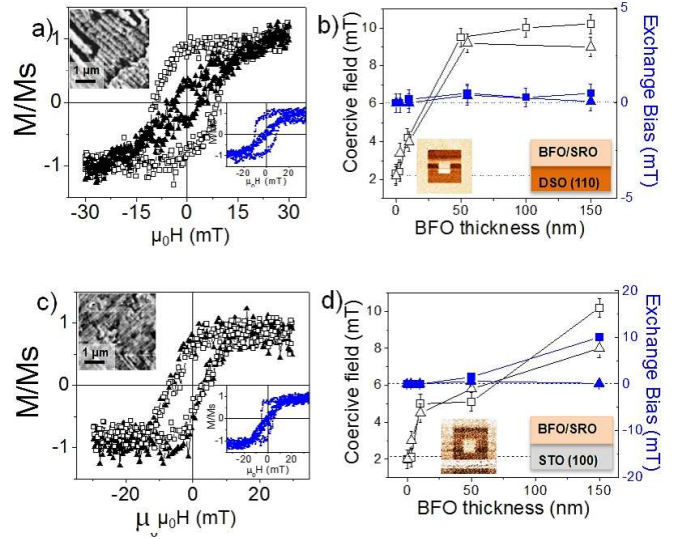}
\caption{\label{fig:fig3} (a) Magnetic in-plane hysteresis loop at room temperature measured on a Pt/CoFe/BFO (50~nm)/SRO/DSO heterostructure, perpendicular to the domain walls (open squares) and along the domain walls (filled triangles) when no growth field was applied. The inset shows the corresponding loop when a growth field was applied and the corresponding MFM image. (b) coercivity enhancement (open symbols, left axes) and exchange bias (closed symbols, right axes) as a function of the multiferroic thickness with (squares) and without (triangles) a growth field. The inset shows the PFM box in a box of the 3~nm thick BFO film. (c-d) Corresponding room temperature loops on a Pt/CoFe/BFO (50~nm)/SRO/STO heterostructure and MFM picture (c) and thickness dependence of the exchange bias field and enhanced coercivity (d).}
\end{figure}

Fig. 3(a) shows the magnetic hysteresis obtained in two orthogonal in-plane orientations from the heterostructure on DSO. A pronounced uniaxial anisotropy is evidenced as in-plane easy and hard axes 90$^{\circ}$ apart as seen in Fig.3 (a). There is also an enhanced coercivity compared to the 2.4~mT coercivity of a CoFe reference layer grown directly on the substrates (not shown). Interestingly, the application of a 20~mT growth field has no effect on this anisotropy set by the two variants BFO multiferroic architecture (see inset in Fig.3(a)), reflecting the strong interfacial coupling.  We note that the 9~mT coercivity value is in the range of the effective coupling field of our simulation. Such macroscopic magnetic behavior is consistent with the multiferroic architecture and the in-plane collinear arrangement of the canted moment and the polarization in BFO schematized in Fig.1(a). In this case of the DSO substrate, the BFO has 71$^{\circ}$ domain walls and there is negligible exchange bias or unidirectional anisotropy.

We then monitored the BFO thickness dependence of the coercive field enhancement and the exchange bias (open and filled symbols, respectively in Fig. 3(b)) along the easy magnetic axis of the CoFe layer with and without a growth field (squares and triangles, respectively). As the BFO thickness decreases from 150~nm to 3~nm, an abrupt change in coercivity can be observed for 10~nm thick films. Below this threshold value, the coercivity remains similar to that for CoFe grown on the substrate instead of BFO. The coercivity saturates for 150~nm BFO films as shown in Fig. 3(b) independently of the application of a magnetic CoFe growth field.

For comparison, the corresponding experiments have been performed for similar heterostructures on STO substrates. For four-fold BFO based heterostructures, no magnetic in-plane anisotropy is observed (Fig. 3(c)). This can be explained by the configuration of the magnetic domains of such films. In inset, the MFM analysis revealed the presence of ferromagnetic stripe domains oriented at 90$^{\circ}$ from each other. Other experiments have shown that the observed exchange bias is correlated with the presence of 109$^{\circ}$ type domain walls, which are present in the four-fold ferroelectric domain structure of BFO films grown on STO,\cite{27,28,29} rather than to a unidirectional anisotropy from the canted antiferromagnetic moments. As the BFO thickness is decreased, the exchange bias field (Fig. 3(d)) disappears before the enhanced coercivity, which decreases abruptly for BFO films thinner than 10~nm. This is in agreement with previous observations\cite{30} as the BFO thin films become single domain and the 109$^{\circ}$ domain walls vanish. In the inset of Fig. 3(b) and (d), the ferroelectric property of the 3~nm BFO thick films was checked. The out-of plane PFM response after the successive application of -3~V and 3~V on both STO and DSO substrates shows that the ferroelectricity of the film is preserved at this thickness. No ferroelectric domains can be resolved in the in-plane PFM response.

\begin{figure}
\centering
\includegraphics[width=0.48\textwidth]{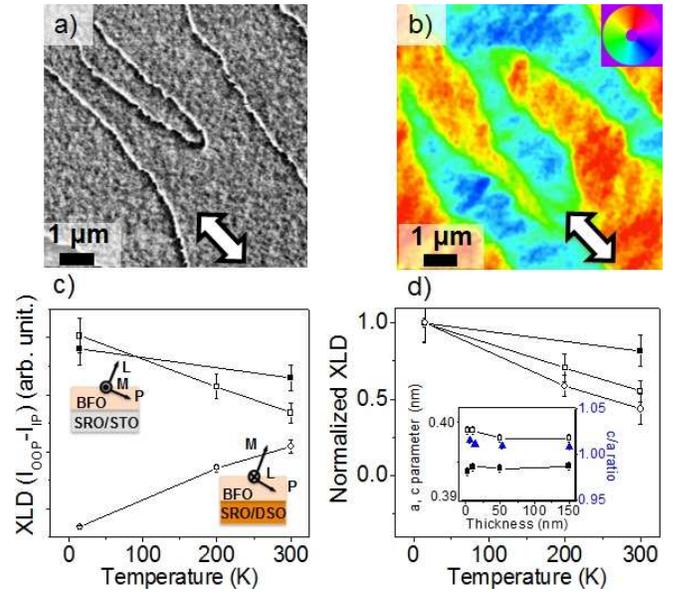}
\caption{\label{fig:fig4} MFM (a) and SEMPA (b) pictures of the CoFe/Pt layer deposited with a growth field (axis labeled by the double arrow) on BFO (3~nm)/SRO/DSO, (c) X-ray linear dichroism as a function of temperature for a 150~nm (black filled squares) and a 3~nm thick (open squares) BFO film on STO and a 3~nm film on DSO (open circles). In inset, schematics of the antiferromagnetic axis $\mathbf L$, polarization vector $\mathbf P$ and canted moment vector $\mathbf M_c$ for the films grown on STO and DSO according to ref.~[\onlinecite{4,5}]. (d) Normalized change in $I_{XLD}$ from room temperature (see text). The inset shows the lattice parameter out of plane $c$ (open squares) and in-plane $a$ (filled squares), as well as the $c/a$ ratio (triangles) as a function of thickness for BFO/SRO/DSO.}
\end{figure}

In order to understand the origin of the apparent reduced influence of the BFO on the CoFe at low thicknesses, MFM and SEMPA measurements were carried out on the 3~nm thick BFO based heterostructures. Figures 4 (a) and (b), respectively, show the MFM and SEMPA images of the CoFe/Pt bilayer on 3~nm BFO/SRO/DSO. Both images show domains oriented along the growth field direction (white filled arrows in Fig. 4. (a) and (b)). The SEMPA image also shows the usual magnetic ripple expected for a CoFe thin film. The correlation between the CoFe magnetization (domains visible in MFM and SEMPA images) and the ferroelectric domains (single domain state from PFM) is lost at 3~nm. At this thickness regime, the growth field sets the magnetic anisotropy, as the influence of the multiferroic BFO becomes negligible.

To further investigate the origin of the decay of the multiferroic influence on the ferromagnet and to get a better understanding of the magnetic ordering in the ultrathin multiferroic layer, X-ray linear dichroism (XLD) experiments were performed at the Fe L-edge in the electron yield mode as function of temperature from 15~K to 300~K. The grazing incident angle was fixed at 30$^{\circ}$, while the photon polarization was rotated by 90$^{\circ}$ to obtain the in-plane and out-of-plane components. The XLD technique probes either the electronic configuration (orbital anisotropy) and/or the long range magnetic order. At room temperature, the ferroelectricity and the antiferromagnetism contribute about equally to the XLD.\cite{31,22} While the ferroelectric order is robust with decreasing thickness in the range studied, as seen by inset in Fig. 3(b and d) and measurements of others,\cite{32,33,34} it is well known that $T_N$ can decrease and hence the antiferromagnetic order is destabilized with decreasing film thicknesses\cite{35}. The magnetic contribution to the XLD intensity is proportional to $|\mathbf m \mathbf L \cdot \mathbf E|$ where $\mathbf E$ is the X-ray electric field and $\mathbf m$ is the Fe magnetic moment, i.e., the antiferromagnetic axis $\mathbf L$.\cite{22,26} The linear dichroism intensity, $I_{XLD}$, is defined as $I_{OOP}-I_{IP}$, the difference in the linearly polarized X-ray absorption (in this case at the Fe L2 edge) with $\mathbf E$ out-of-plane and in-plane. $I_{XLD}$ is plotted as a function of temperature in Fig. 4(c) for 150~nm and 3~nm BFO films on STO and for a 3~nm BFO film on DSO. The difference in the sign of the slope on the two different substrates can be explained by strain considerations. Depending on the BFO strain state, the antiferromagnetic axis can be oriented along $<112>$ with an out-of-plane component as in the case of high compressive in-plane strain on STO,\cite{4} or fully in-plane along $<110>$ for lower in-plane strain on DSO,\cite{5} as seen in insets of Fig. 4 (c). On STO the antiferromagnetic axis makes a 54$^{\circ}$ angle with the sample surface and the $I_{OOP}$ dominates $I_{XLD}$, which is consistent with the observed increasing dichroism with increasing antiferromagnetic order at low temperature. On the DSO substrate, the $I_{XLD}$ is mainly due to the in-plane component $I_{IP}$, and $I_{XLD}$ decreases with decreasing temperature. For thick films, the data is taken far from both the ferroelectric Curie temperature, $T_C =1143$~K, and the N\'eel temperature, $T_N = 673$~K\cite{37} and little change as a function of temperature in the XLD is expected in agreement with our observation (black filled squares in Fig. 4(c)). However, for the thinner films, changes can occur due to the changing antiferromagnetic order. Significant temperature dependence is observed for 3~nm thick BFO films on both DSO and STO substrates, open circles and open squares, respectively.

The temperature dependence of the antiferromagnetic order is more easily seen by replotting in Fig. 4(d) the data of Fig. 4(c) to show the normalized magnitude of the relative temperature dependence below room temperature, $[|I_{XLD}(T) - I_{XLD}(300~\text{K})|+I_{XLD}(300~\text{K})]/[| I_{XLD}(15~\text{K}) - I_{XLD}(300~\text{K})| + I_{XLD}(300~\text{K})]$. For the 3~nm thick BFO films, a decrease to approximately 50\% is observed at room temperature compared to the bulk value which is recovered at 15~K. Considering the nearly constant tetragonality of the BFO thin films over the considered thickness range, as shown in inset of Fig. 4(d) and from previous work,\cite{32} it is then reasonable to assume that the decrease in the coupling strength we observe with decreasing BFO thickness is not strain induced. Rather it can be attributed to a reduced T$_N$ and hence to a reduced antiferromagnetic order.

\section{Summary}

In summary, we have investigated the nature of the coupling between the multiferroic BFO and the CoFe magnetic film. Measurements as a function of BFO thickness show that the coupling, resulting in enhanced CoFe coercivity disappears when the antiferromagnetism disappears. Although low BFO thickness is desirable to minimize switching voltages in devices, a lower thickness limit of at least 10~nm is established for adequate coupling. High resolution imaging of the magnetization shows that, within the constraints of the intralayer exchange coupling in the CoFe film, the magnetization follows the in-plane polarization in a ferroelectric domain. The magnetization images along with no significant exchange bias indicate uniaxial rather than unidirectional anisotropy in the CoFe/BFO coupling. The behavior is consistent with either an interfacial exchange coupling between the CoFe moments and a canted antiferromagnetic moment in BFO (e.g.\ spin-flop coupling), or coupling to a DM induced canted moment\cite{13} with a weak barrier to rotation of the oxygen octahedra. An interfacial coupling field of order 10~mT is derived from the micromagnetic modeling and the coercivity enhancement of coupled BFO/CoFe structures.

We sincerely thank D.~Meier and R.~Ramesh for the thoughtful discussions and B.~Kim and Y.~Bobrov for their assistance with the MFM under external magnetic field. We also gratefully acknowledge helpful discussions with R.~D.~McMichael and M.~D.~Stiles and the technical assistance of S.~Blankenship and G.~Holland. S.~Bowden acknowledges support under the Cooperative Research Agreement between the University of Maryland and the National Institute of Standards and Technology Center for Nanoscale Science and Technology, Award 70NANB10H193, through the University of Maryland. M.~Trassin acknowledges the support from the Center for Energy Efficient Electronics Science (NSF Award 0939514).

$^*$Email address: morgan.trassin@mat.ethz.ch


\end{document}